\documentclass[%
preprint,
 amsmath,amssymb,
 aps,
 superscriptaddress
]{revtex4-1}
\usepackage[dvipdfmx]{graphicx}% Include figure files
\usepackage{bm}
\usepackage[top=20truemm,bottom=20truemm,left=20truemm,right=20truemm]{geometry}
\usepackage{array}
\usepackage{color}
\usepackage{setspace}
\usepackage{mathtools}

\begin{document}
\setlength\textfloatsep{11pt}
%\preprint{APS/123-QED}

\title{Magnetization process of the insulating ferromagnetic semiconductor (Al,Fe)Sb}% Force line breaks with \\

\author{Shoya Sakamoto}
\affiliation{Department of Physics, The University of Tokyo, Bunkyo-ku, Tokyo 113-0033, Japan}%Lines break automatically or can be forced with \\

\author{Le Duc Anh}%
\affiliation{Department of Electrical Engineering and Information Systems, The University of Tokyo, Bunkyo-ku, Tokyo 113-8656, Japan}
\affiliation{Institute of Engineering Innovation, The University of Tokyo, Bunkyo-ku, Tokyo 133-8656, Japan}

\author{Pham Nam Hai}%
\affiliation{Department of Electrical Engineering and Information Systems, The University of Tokyo, Bunkyo-ku, Tokyo 113-8656, Japan}
\affiliation{Department of Electrical and Electronic Engineering, Tokyo Institute of Technology, Meguro-ku, Tokyo 152-0033, Japan}
\affiliation{Center for Spintronics Research Network (CSRN), The University of Tokyo, Bunkyo-ku, Tokyo 113-8656, Japan}

\author{Yukiharu Takeda}
\affiliation{Materials Sciences Research Center, Japan Atomic Energy Agency (JAEA), Sayo-gun, Hyogo 679-5148, Japan}

\author{Masaki Kobayashi}

\affiliation{Department of Electrical Engineering and Information Systems, The University of Tokyo, Bunkyo-ku, Tokyo 113-8656, Japan}
\affiliation{Center for Spintronics Research Network (CSRN), The University of Tokyo, Bunkyo-ku, Tokyo 113-8656, Japan}

\author{Yuki K. Wakabayashi}
\affiliation{Department of Electrical Engineering and Information Systems, The University of Tokyo, Bunkyo-ku, Tokyo 113-8656, Japan}

\author{Yosuke Nonaka}
\affiliation{Department of Physics, The University of Tokyo, Bunkyo-ku, Tokyo 113-0033, Japan}

\author{Keisuke Ikeda}
\affiliation{Department of Physics, The University of Tokyo, Bunkyo-ku, Tokyo 113-0033, Japan}

\author{Zhendong Chi}
\affiliation{Department of Physics, The University of Tokyo, Bunkyo-ku, Tokyo 113-0033, Japan}

\author{Yuxuan Wan}
\affiliation{Department of Physics, The University of Tokyo, Bunkyo-ku, Tokyo 113-0033, Japan}

\author{Masahiro Suzuki}
\affiliation{Department of Physics, The University of Tokyo, Bunkyo-ku, Tokyo 113-0033, Japan}

\author{Yuji Saitoh}
\affiliation{Materials Sciences Research Center, Japan Atomic Energy Agency (JAEA), Sayo-gun, Hyogo 679-5148, Japan}

\author{Hiroshi Yamagami}
\affiliation{Materials Sciences Research Center, Japan Atomic Energy Agency (JAEA), Sayo-gun, Hyogo 679-5148, Japan}
\affiliation{Department of Physics, Kyoto Sangyo University, Kyoto 603-8555, Japan}

\author{Masaaki Tanaka}
\affiliation{Department of Electrical Engineering and Information Systems, The University of Tokyo, Bunkyo-ku, Tokyo 113-8656, Japan}
\affiliation{Center for Spintronics Research Network (CSRN), The University of Tokyo, Bunkyo-ku, Tokyo 113-8656, Japan}

\author{Atsushi Fujimori}
\affiliation{Department of Physics, The University of Tokyo, Bunkyo-ku, Tokyo 113-0033, Japan}



\date{\today}% It is always \today, today,

\begin{abstract}

We have studied the magnetization process of the new insulating ferromagnetic semiconductor (Al,Fe)Sb by means of x-ray magnetic circular dichroism.
For an optimally doped sample with 10\% Fe, a magnetization was found to rapidly increase at low magnetic fields and to saturate at high magnetic fields at room temperature, well above the Curie temperature of 40 K. 
We attribute this behavior to the existence of nanoscale Fe-rich ferromagnetic domains acting as superparamagnets.
By fitting the magnetization curves using the Langevin function representing superparamagnetism plus the paramagnetic linear function, we estimated the average magnetic moment of the nanoscale ferromagnetic domain to be $300$-$400$$\mu_{B}$, and the fraction of Fe atoms participating in the nano-scale ferromagnetism to be $\sim$50\%.
Such behavior was also reported for (In,Fe)As:Be and Ge:Fe, and seems to be a universal characteristic of the Fe-doped ferromagnetic semiconductors.
Further Fe doping up to 14\% led to the weakening of the ferromagnetism probably because antiferromagnetic superexchange interaction between nearest-neighbor Fe-Fe pairs becomes dominant.

\end{abstract}

\pacs{Valid PACS appear here}% PACS, the Physics and Astronomy
\maketitle

%\tableofcontents

\section{Introduction}

Recently, Fe-doped ferromagnetic semiconductors (FMS) were discovered and have attracted much attention owing to their high Curie temperatures ($T_{\rm C}$) and distinct properties compared to the prototypical Mn-doped systems.
First, the $T_{C}$'s are higher than room temperature: 340 K for (Ga,Fe)Sb \cite{Tu:2016aa} and 335 K for (In,Fe)Sb \cite{Tu:2018aa}, while that of (Ga,Mn)As is at most 200 K \cite{Chen:2011aa}. 
Second, various types of transport properties are realized: $p$-type for (Ga,Fe)Sb \cite{Tu:2015aa} and Ge:Fe \cite{Wakabayashi:2014ab}, $n$-type for (In,Fe)As \cite{Nam-Hai:2012ac} and (In,Fe)Sb \cite{Tu:2018aa}, and insulating for (Al,Fe)Sb \cite{Anh:2015aa}, while only $p$-type is possible for Mn-doped III-V group semiconductors.
Furthermore, in the case of (In,Fe)As, a large Curie temperature modulation of 42\% was demonstrated by wave function engineering in field-effect transistor (FET) structures \cite{Anh:2015ab}, and the spin splitting of the conduction band bottom was observed in Esaki-diode structures \cite{Le-Duc-Anh:2016aa}.
Despite the attractive properties, however, the origin of ferromagnetism is unclear and remains to be investigated.

%In the present study, we have studied the new insulating FMS (Al,Fe)Sb. 

In the insulating FMS (Al,Fe)Sb, ferromagnetism of $T_{\rm C}$ = 40 K emerges for 10\% of Fe doping, and the hole concentration is about 10$^{17}$ cm$^{-3}$ \cite{Anh:2015aa}.
Unlike the other FMSs, further Fe doping up to 14\% leads to the decrease in $T_{\rm C}$ down to 10 K and the two orders of magnitudes increase in the hole concentration up to 10$^{19}$ cm$^{-3}$.
This is probably because the crystal quality is degraded through the creation of high concentration of defects, although there is no clear signature of phase separation from reflection high-energy electron-diffraction (RHEED) patterns and x-ray diffraction (XRD) profiles.

Since (Al,Fe)Sb is insulating, the ferromagnetism is most likely not carrier-induced being different from the other typical FMSs \cite{Dietl:2014aa}. 
In the present study, for the purpose of revealing the mechanism of how ferromagnetism appears in (Al,Fe)Sb, we have conducted x-ray absorption spectroscopy (XAS) and x-ray magnetic circular dichroism (XMCD) measurements at the Fe $L_{2,3}$ absorption edges.
XAS and XMCD are powerful methods to investigate the magnetism as well as the local electronic structure of specific elements and have been used for the studies of magnetic semiconductors \cite{Edmonds:2005aa,Takeda:2008aa}.
Since XAS and XMCD are element specific probes, one can eliminate the diamagnetic component from the substrate, which is usually dominant in the case of diluted magnetic materials in thin film form. One can thus precisely estimate the paramagnetic (PM) linear component as well as the ferromagnetic (FM) component from magnetization curves \cite{Sakamoto:2016aa}.

In the previous XMCD study on (In,Fe)As:Be \cite{Sakamoto:2016aa} and Ge:Fe \cite{Wakabayashi:2016aa}, it was found that nano-scale FM domains exist far above the Curie temperature. On lowering the temperature, those domains seemed to coalesce, resulting in global ferromagnetism at the Curie temperature.
In the present study, we have found the same behavior for (Al,Fe)Sb and concluded that the nanoscale FM domains of several hundreds $\mu_{\rm B}$ are formed in Fe-rich regions, which is likely the universal feature of Fe-doped magnetic semiconductors.

\section{experiment}
Two samples Al$_{0.9}$Fe$_{0.1}$Sb and Al$_{0.86}$Fe$_{0.14}$Sb were grown using the low-temperature molecular beam epitaxy (LT-MBE) methods. The structure of the samples was, from the top surface to the bottom, InAs cap (5nm)/(Al,Fe)Sb (100 nm)/AlSb (100 nm)/GaAs (100 nm)/p-GaAs(001) substrate. The detail of the sample growth is described in Ref. \cite{Anh:2015aa}.

The XAS and XMCD experiments were conducted at beam line BL23SU of SPring-8. 
Measurement temperature $T$ was varied from 5.8 K to 300 K, and magnetic field $\mu_{0}H$ from -7 T to 7 T. 
The samples were placed in the measurement chamber so that the sample surface was perpendicular to the x-ray incident direction and hence the magnetic field.
Absorption signals were collected in the total electron yield (TEY) mode, and dichroic signals were measured by reversing the helicity of x rays with 1 Hz frequency at each photon energy. % under a fixed magnetic field. 
The spectra were obtained by sweeping the photon energy under a fixed magnetic field, and the scans were repeated with the opposite magnetic field direction to minimize experimental artifacts. 
That is, each XMCD spectrum was obtained as $(\sigma_{+,h}-\sigma_{-,h})+(\sigma_{-,-h}-\sigma_{+,-h})$, and each XAS spectrum as the summation of all the four terms, where $\sigma$ denotes the absorption cross-sections, the first subscript the helicity of x rays, and the second subscript the direction of magnetic field. 
We also measured magnetization curves by recording XMCD signals at the photon energies of the $L_{2,3}$ edges while sweeping the magnetic field. The data have been normalized to the total magnetic moment at $\mu_{0}H$ = 7 T deduced by applying the XMCD sum rules to the spectra \cite{Thole:1992aa,Carra:1993aa,Chen:1995aa}. 

\section{Results and Discussion}

\begin{figure}
\begin{center}
\includegraphics[width=8cm]{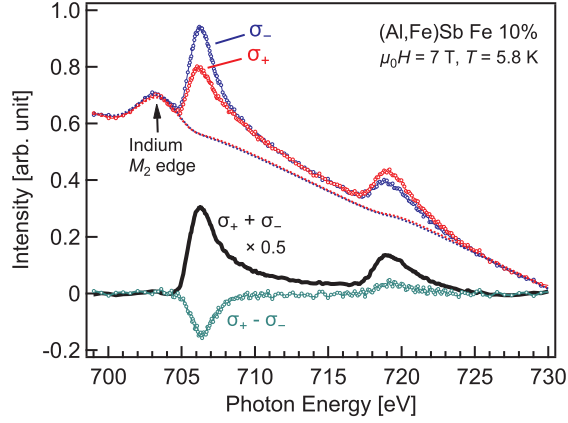}
\caption{Absorption spectra taken with x rays of positive and negative helicities. Dashed curves at the top figure represent the backgrounds, which consist of a Lorentzian function representing the Indium $M_{2}$ peak, a double-step function for edge jumps, and a linear function. At the bottom, XAS and XMCD spectra after the background subtraction are also shown.}
\label{AlFeSb_XAS_BG}
\end{center}
\end{figure}

Figure \ref{AlFeSb_XAS_BG} shows raw absorption spectra taken with x rays of positive and negative helicities under $\mu_{0}H$ = 7 T and $T$ = 5.8 K. A large difference between the spectra taken with different helicities indicates that the magnetism indeed arises from the Fe atoms.
Because there was a 5-nm-thick InAs capping layer to prevent oxidation, a relatively strong In $M_{2}$ peak overlapped the Fe $L_{3}$ peak. 
In order to remove the In contribution, we assumed a Lorentzian function and subtracted it from the spectra together with a linear background. Double step functions representing the Fe $L_{2,3}$-edge jumps have also been subtracted.
The summation and the difference spectra of the different helicities after the background subtraction are shown at the bottom of Fig. \ref{AlFeSb_XAS_BG}. 
Note that we have processed all the data in the same manner.

\begin{figure}
\begin{center}
\includegraphics[width=8.4cm]{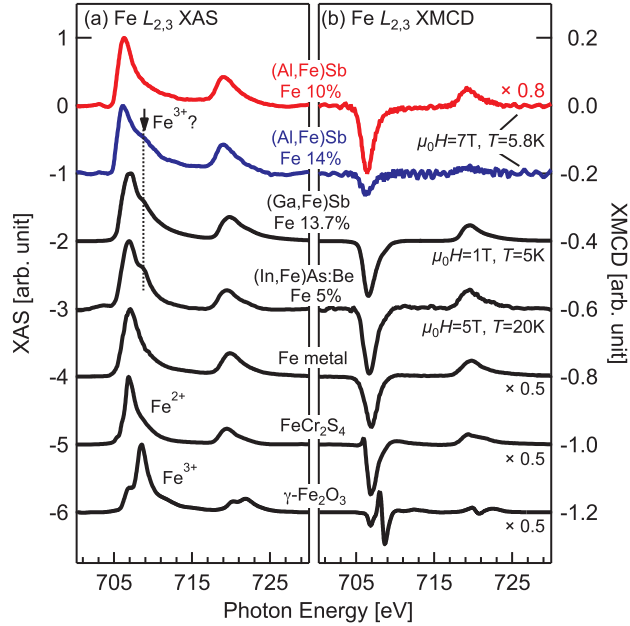}
\caption{XAS and XMCD spectra of (Al,Fe)Sb in comparison with those of (Ga,Fe)Sb \cite{Verma_FeCr2S4}, (In,Fe)As:Be \cite{Sakamoto:2016aa}, Fe metal \cite{Chen:1995aa}, FeCr$_{2}$S$_{4}$, and $\gamma$-Fe$_{2}$O$_{3}$ \cite{Brice-Profeta:2005aa}.}
\label{AlFeSb_XAS_XMCD}
\end{center}
\end{figure}

Figure \ref{AlFeSb_XAS_XMCD} shows the averaged XAS and XMCD spectra of (Al,Fe)Sb taken under two opposite magnetic-field directions as described in Section II.
The spectra of the other Fe-doped FMSs, namely, (Ga,Fe)Sb and (In,Fe)As:Be as well as those of Fe metal, FeCr$_{2}$S$_{4}$ (Fe$^{2+}$), and Fe$_{2}$O$_{3}$ (Fe$^{3+}$) are shown for comparison.
The XAS and XMCD spectra of (Al,Fe)Sb are broad and asymmetric having a tail on the high-energy side, similar to the case of Fe metal.
In addition, fine structures due to multiplet splitting are hardly seen, which would be present if the Fe 3$d$ electrons are localized as in Fe$_{2}$O$_{3}$ and FeCr$_{2}$S$_{4}$.
These indicate the significant delocalization of the Fe 3$d$ electrons and would challenge the assumption that Fe takes the valence of 3+ with localized five 3$d$ electrons.
The same is also true for the other Fe-doped FMSs \cite{Sakamoto:2016aa, Wakabayashi:2016aa}, and the itinerancy of Fe 3$d$ electrons seems to be a key for the ferromagnetism of the Fe-doped systems.
Furthermore, the XMCD spectra of (Al,Fe)Sb, (Ga,Fe)Sb, and (In,Fe)As:Be have almost identical line shapes, indicating that the local electronic structure of the Fe atoms that contribute to the ferromagnetism is similar among the Fe-doped FMSs.
Note that the shoulders around 709 eV that the XAS spectra of (Ga,Fe)Sb and (In,Fe)As:Be have were attributed to Fe$^{3+}$ signals originating from Fe oxides formed near the sample surfaces \cite{Sakamoto:2016aa, Sakamoto:2018aa} because a corresponding feature does not exist in the XMCD spectra.
The XAS spectrum of the 14\% Fe-doped AlSb also exhibits such a shoulder, which may be either due to the existence of surface oxides as in the cases of (Ga,Fe)Sb and (In,Fe)As:Be or due to some defect states considering the lower crystal quality of the 14\% Fe-doped AlSb sample lower than the 10\% Fe-doped one.

The XMCD intensity is considerably weaker for the 14\% Fe-doped than for the 10\% Fe-doped sample, being consistent with the observation that the Curie temperature decreases with increasing Fe concentration from 10\% to 14\%.
This also suggests that the ferromagnetism of (Al,Fe)Sb is not due to Fe-metal precipitates or any secondary phases because if the magnetism originated from Fe-metal precipitates, higher Fe doping level would result in enhanced ferromagnetism. 
Considering that the spectral line shapes of XAS and XMCD are similar between the two samples except for the smaller shoulder in the XAS spectrum of 10\% Fe-doped sample, the disappearance or the weakening of the ferromagnetism with increasing Fe concentration seems also intrinsic.

The spin and orbital magnetic moments at $\mu_{0}H$ = 7 T and $T$ = 5.8 K estimated from the XMCD sum rules were ($m_{\rm spin}$, $m_{\rm orb}$) = (1.63$\mu_{\rm B}$, 0.423$\mu_{\rm B}$) for the 10\% Fe-doped sample and (0.38$\mu_{\rm B}$, 0.08$\mu_{\rm B}$) for the 14\% Fe-doped one. Here, we have assumed the number of 3$d$ electrons to be 6, which was deduced from first-principles supercell calculations \cite{sakamoto_DFT}, and the correction factor to be 0.875 for the Fe$^{2+}$ state \cite{Teramura:1996aa, Piamonteze:2009aa}.
The unquenched orbital magnetic moments seem to support the mixture of the $3d^6\underline{L}$ configuration of the Fe atom, where $\underline{L}$ denotes a ligand hole.
Note that the uncertainty in $m_{\rm orb}/m_{\rm spin}$ and $m_{\rm orb}+m_{\rm spin}$ can be as large as $\sim20\%$ and $\sim10\%$, respectively, because the thick InAs capping layer made signals very weak.

\begin{figure}
\begin{center}
\includegraphics[width=8.5cm]{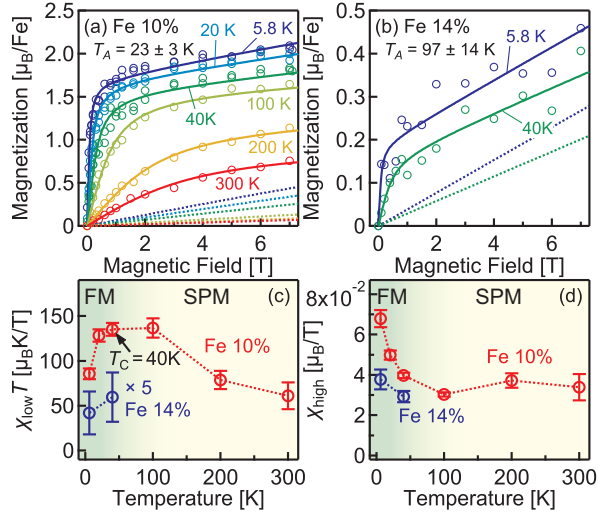}
\caption{(a) Magnetization curves of the 10\% Fe-doped sample and (b) the 14\% Fe-doped sample deduced from XMCD. The fitting results are shown by solid curves, and the linear PM components are also shown separately by dashed lines. (c) Magnetic susceptibility at low magnetic fields near 0 T $\chi_{\rm low}$ [defined by Eq. (\ref{eq:chi})] and (d) that at high magnetic fields around 7 T $\chi_{\rm high}$ [also defined by Eq. (\ref{eq:chi})], which have been deduced from the fitting. Here,  $\chi_{\rm low}$ is multiplied by temperature.}
\label{AlFeSb_MH}
\end{center}
\end{figure}

Figures \ref{AlFeSb_MH}(a) and \ref{AlFeSb_MH}(b) show the magnetization curves at various temperatures measured by XMCD. 
Here, hysteresis could not be detected because it was too small.
%The magnetizations exhibit steep increase at low magnetic fields and gradual linear increase at high magnetic fields. 
Despite the fact that the $T_{\rm C}$ below which hysteresis appears is as low as 40 K for the 10\% Fe-doped sample, the relatively large magnetization $M$ is induced by the magnetic field even at 300 K in contrast to typical ferromagnets, where magnetization disappears rather quickly above $T_{C}$. 
We attribute this behavior to the superparamagnetism of nanoscale FM domains, as has been found in other Fe-doped FMSs \cite{Sakamoto:2016aa,Wakabayashi:2016aa}.
At the low temperature of 5.8 K, the magnetization at high magnetic fields shows a gradual linear increase after the steep increase at low magnetic fields.
This indicates the existence of PM Fe atoms even below $T_{\rm C}$.

To be quantitative, the magnetic susceptibility or the slope of the magnetization versus magnetic field curve at low and high magnetic fields,
\begin{equation}
\chi_{\rm low}=\left.\frac{\Delta M}{\Delta (\mu_{0}H)}\right|_{\mu_{0}H = 0 {\rm T}}, \left.\chi_{\rm high}=\frac{\Delta M}{\Delta (\mu_{0}H)}\right|_{\mu_{0}H = 7 {\rm T}},
\label{eq:chi}
\end{equation} 
are plotted in Figs. \ref{AlFeSb_MH}(c) and \ref{AlFeSb_MH}(d), respectively.
Note that, in Fig. \ref{AlFeSb_MH}(c), $\chi_{\rm low}$ is multiplied by the temperature to see whether $\chi_{\rm low}$ is inversely proportional to temperature as in ideal superparamagnets.  
Although the $\chi_{\rm low}T$ of the 10\% Fe-doped sample shows some temperature dependence, it remains large even at 300 K. This again indicates the existence of superparamagnetism. On lowering the temperature, $\chi_{\rm low}T$ gradually increases probably because the number of Fe atoms participating in superparamagnetism increases. Further lowering the temperature down to below $T_{\rm C}$ = 40 K leads to a drop of $\chi_{\rm low}T$. This implies that the system turns into global ferromagnetism which cannot be explained by superparamagnetism. 

%Although $\chi_{\rm low}$ of the sample with 14\% Fe is much smaller than that of the sample with 10\%, $\chi_{\rm high}$ is comparable.
The finite $\chi_{\rm high}$ at the low temperature of 5.8 K indicates the coexistence of ferromagnetism and paramagnetism as mentioned above. 
The values are comparable between the two samples although $\chi_{\rm high}$ of the sample with 14\% Fe could be much larger than that of the sample with 10\% Fe considering that there are much more PM Fe atoms.
This behavior can be understood if a large fraction of Fe atoms are coupled antiferromagnetically in the 14\% Fe-doped sample.

In order to understand the behavior of $\chi_{\rm low}$ and $\chi_{\rm high}$ or how the PM and superparamagnetic (SPM)/FM phase coexist in more detail, we fitted the linear combination of the Langevin function $L(\xi)$ representing superparamagnetism and a linear function representing paramagnetism to the data:
\begin{eqnarray}
\label{eq:fit}
&\displaystyle M = xm_{\rm sat}L(\frac{\mu\mu_{0}H}{k_{\rm B}T}) + (1-x)\frac{C\mu_{0}H}{T + T_{A}},\\
\label{eq:curie}
&\displaystyle C = \frac{m_{\rm sat}(m_{\rm sat}+2\mu_{\rm B})}{3k_{\rm B}},\\
\label{eq:Langevin}
&\displaystyle L(\xi) = \coth(\xi)-\frac{1}{\xi},
\end{eqnarray}
where $M$ is the magnetization per Fe atom, $m_{\rm sat}$ the total magnetic moment of Fe atom, $k_{B}$ the Boltzmann constant, and $C$ the Curie constant. We have assumed that $m_{\rm sat} = 3.4\mu_{\rm B}$, which was obtained by first-principles supercell calculation, and that the $g$ factor is 2 for simplicity.
Fitting parameters in the present model are the following: $\mu$, the total magnetic moment of a nanoscale FM domain; $x$, the fraction of Fe atoms participating in ferromagnetism or superparamagnetism; $T_{A}$, the antiferromagnetic Weiss temperature. Note that $\mu$ and $x$ were allowed to vary with temperature, while $T_{A}$ was kept constant.

The fitted curves are shown by solid curves and the linear PM components are separately shown by dashed lines in Figs. \ref{AlFeSb_MH}(a) and \ref{AlFeSb_MH}(b). As can be seen, the data were fitted well by the Eqs. (\ref{eq:fit})-(\ref{eq:Langevin}).
The fit yielded the antiferromagnetic Weiss temperature of 23 K for the sample with 10\% Fe and 97 K for the sample with 14\% Fe, again indicating that antiferromagnetic correlations are stronger for the sample with 14\% Fe.
This may be because antiferromagnetic superexchange interactions between adjacent Fe atoms become dominant when the concentration of doped Fe atoms increases.
Note that using the Brillouin function instead of a linear function does not change the line shape of the fitting curves as well as the fitting parameters, because relatively large Weiss temperatures make the Brillouin function linear within the range of 0 T to 7 T even at the lowest temperature of 5.8 K.

\begin{figure}
\begin{center}
\includegraphics[width=8.5cm]{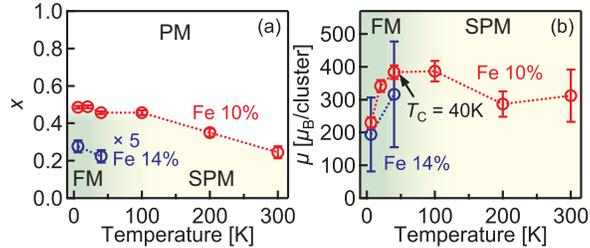}
\caption{Fitting parameters. (a) Fraction of Fe atoms contributing to ferromagnetism or superparamagnetism, denoted by $x$. (b) Total magnetic moment $\mu$ per nanoscale FM domain.}
\label{AlFeSb_parameters}
\end{center}
\end{figure}

Figures \ref{AlFeSb_parameters}(a) and \ref{AlFeSb_parameters}(b) show the temperature dependences of the obtained fitting parameters $x$, the fraction of Fe atoms participating in ferromagnetism or superparamagnetism, and $\mu$, the total magnetic moment of each nanoscale FM domain, respectively. 
$x$ for the sample with 14\% Fe was about only 5\%, which confirms the near absence of ferromagnetism in the sample with 14\% Fe.
In the case of the 10\% Fe-doped sample, 25\% of Fe atoms still contribute to the ferromagnetism or superparamagnetism at 300 K. With decreasing temperature, $x$ gradually increases, while 50\% Fe remain PM at 5.8 K.
These observations highlight the inhomogeneous nature of magnetism in (Al,Fe)Sb. %From the temperature dependence of $x$, it can be inferred that the $T_{C}$ of FM/SPM domains is very high.
The magnetic moment per nanoscale FM domain was found to be 300-400$\mu_{\rm B}$ and increased from $\sim$300$\mu_{\rm B}$ to $\sim$400$\mu_{\rm B}$ with decreasing temperature from 300 K to 40 K. Note that $\mu$ dropped suddenly below 40 K, as $\chi_{\rm low}T$ dropped, reflecting the gradual appearance of macroscopic ferromagnetism and concomitant deviation from the Langevin behavior.
Such a SPM response from the nanoscale FM domains with several hundreds of $\mu_{\rm B}$ was also reported for (In,Fe)As:Be \cite{Sakamoto:2016aa} and Ge:Fe \cite{Wakabayashi:2016aa} and seems to be a universal feature of the Fe-doped FMSs.
Here, $\mu$ of 300$\mu_{\rm B}$-400$\mu_{\rm B}$ corresponds to $\mu/m_{\rm Fe} \sim$ 100 Fe atoms. 
The density of nanoscale FM domains could be deduced from $\mu$ and $x$ as $7 \times 10^{18}$ cm$^{-3}$ for the10\% Fe-doped sample and $1 \times 10^{18}$ cm$^{-3}$ for the 14\% Fe-doped sample. 

%Since the ferromagnetic interaction is strong persisting at least up to room temperature, 
The origin of the nanoscale FM domains can be attributed to the nanoscale fluctuation of Fe distribution as discussed for the other Fe-doped semiconductors (In,Fe)As:Be \cite{Sakamoto:2016aa, Vu:2014aa, Yuan:2018aa} and Ge:Fe \cite{Wakabayashi:2016aa, Shinya:2017aa, Sakamoto:2017aa}. In addition, the previous theoretical studies where chemical pair interactions were calculated suggested that Fe atoms tend to segregate and form Fe-rich regions in the InAs \cite{Vu:2014aa} and Ge \cite{Shinya:2017aa} matrices, while maintaining the zinc blende and diamond lattice structures.
%Note that in the case of (In,Fe)As:Be, the formation of first nearest-neighbor Fe-Fe pairs is not favorable energetically \textcolor{red}{but second nearest-neighbor Fe-Fe pairs is \cite{Vu:2014aa}}.
The same kind of scenario is likely to apply to (Al,Fe)Sb and possibly to the other Fe-doped FMSs. 
If so, FM interaction between Fe 3$d$ orbitals can stabilize the ferromagnetism in nanoscale Fe-rich FM domains at rather high temperatures up to 300 K.
This is consistent with the present XAS and XMCD spectra, where rather close Fe-Fe distance bestows the itinerancy of the 3$d$ electrons.
On lowering the temperature, those domains would start to overlap or interact with each other, resulting in the macroscopic ferromagnetism.

%As mentioned above, the magnetization of the sample with 14\% Fe is much smaller than that of the sample with 10\% Fe. 
%Moreover, it is mostly composed of linear paramagnetic-like component meaning that the ferromagnetism is almost absent in the sample with 14\% Fe.

\section{Summary}
We have performed x-ray absorption spectroscopy (XAS) and x-ray magnetic circular dichroism (XMCD) on the new ferromagnetic semiconductor (Al,Fe)Sb to study its electronic structure and magnetization process.
The spectral line shapes were broad and asymmetric having a tail on the high-energy side, indicating the itinerant nature of Fe 3$d$ electrons. 
The XMCD sum rules yielded an unquenched orbital magnetic moment, which suggests that a considerable fraction of Fe atoms take the 3$d^{6}$ configuration with a ligand hole. 
From the magnetization curves measured by XMCD, we have found that nanoscale ferromagnetic domains of 300-400 $\mu_{\rm B}$ exist even at room temperature in the optimally doped sample with 10\% Fe ($T_{\rm C}=40$ K), and the system behaves as superparamagnets.
The formation of such domains was ascribed to the non-uniform distribution of Fe atoms on the nanoscale, which appears to be a common characteristics of the other Fe-doped FMSs.
For the 14\% Fe-doped sample, the weakening of the ferromagnetism and the strengthening of the antiferromagnetic correlations were observed compared to the 10\% Fe-doped sample. 
This may be because antiferromagnetic superexchange interaction between adjacent Fe atoms becomes dominant when the system is doped with a large concentration of Fe atoms.

\section*{acknowledgments}
This work was supported by Grants-in-Aids for Scientific Research from JSPS (No. 15H02109 and 17H04922).
The experiment was done under the Shared Use Program of JAEA Facilities (Proposal Nos. 2016B-E20 and 2017A-E20) with the approval of the Nanotechnology Platform Project supported by MEXT. The synchrotron radiation experiments were performed at the JAEA beamline BL23SU in SPring-8 (Proposal Nos. 2016B3841 and 2017A3841).
A.F. is an adjunct member of the Center for Spintronics Research Network (CSRN), the University of Tokyo, under Spintronics Research Network of Japan (Spin-RNJ).
S.S. acknowledges financial support from the Advanced Leading Graduate Course for Photon Science (ALPS) and the JSPS Research Fellowship for Young Scientists. Z.C. acknowledges financial support from the Materials Education Program for the Futures Leaders in Research, Industry and Technology (MERIT). 

\bibliography{AlFeSb_XMCD2.bib}

\end{document}